\documentclass{article}

\usepackage{amssymb,latexsym}

\usepackage[pdftex]{graphicx}

\usepackage{hyperref}

\begin{document}

\pagestyle{plain}

\newtheorem{theorem}{Theorem}[section]

\newtheorem{proposition}[theorem]{Proposition}

\newtheorem{lemma}[theorem]{Lemma}

\newtheorem{corollary}[theorem]{Corollary}

\newtheorem{definition}[theorem]{Definition}

\newtheorem{remark}[theorem]{Remark}

\newtheorem{exempl}{Example}[section]

\newenvironment{example}{\begin{exempl}  \em}{\hfill $\square$

\end{exempl}}  \vspace{.5cm}

\renewcommand{\contentsname}{ }

\title{Molecular computers}

\author{Marius Buliga \\ 
\\
Institute of Mathematics, Romanian Academy \\
P.O. BOX 1-764, RO 014700\\
Bucure\c sti, Romania\\ 
{\footnotesize Marius.Buliga@imar.ro , mbuliga@protonmail.ch}}  \vspace{.5cm}

\date{(2015)}

\maketitle

\begin{abstract}
We propose the chemlambda artificial chemistry, whose behavior strongly suggests that real molecules which embed Interaction Nets patterns and real chemical reactions which resemble Interaction Nets graph rewrites could be a realistic path towards molecular computers, in the sense explained in the article. 
\end{abstract}

\paragraph{Note to the reader.} This is a text version of the article \cite{0}, with only minimal edits. The original article was part of an experiment in Open Science. The goal was to report research which can be independently validated in the most simple way possible. It is part of a Github repository \cite{4}. The original article uses javascript animations which are created with the programs from the repository. The versions history of the original article is available at \href{https://github.com/chorasimilarity/chemlambda-gui/commits/gh-pages/dynamic/molecular.html}{this link}. In the same repository there is a library of 427 "molecules" (graphs) \cite{8} and demos pages \cite{9}. Many simulations results are available at \cite{10}. Everything from the article and the other cited sources, excepting the suggestion that such molecular computers can be done with real chemicals, can be validated by using the resources provided. 

The particular formatting of the article is tributary to the original \cite{0}. Animations which are embedded in the original are replaced with links. The references and links in the text are left as they were and much improvement could be made  concerning references pointing to Interaction Nets \cite{12}, for example; Berry and Boudol CHAM \cite{13} is not explicitely mentioned here, but the initial version of this formalism, i.e. "chemical concrete machine"\cite{2},  uses it. The Algorithmic Chemistry of Fontana and Buss \cite{14} \cite{15} appears here via a link to their research, perhaps \href{http://tuvalu.santafe.edu/~walter/AlChemy/alchemy.html}{this link} is  better.  The same  happens with the work \cite{16} by Christoph Flamm, for which, in the body of the article, a link to his papers page is provided.  Part of this was due to the ignorance of the author  at the moment (a mathematician). A later hope was to keep the article as is and to link it with critical public reviews, as a weaker form of validation than, for example, an alternative Haskell implementation of chemlambda \cite{16}.   However, the main idea of the article is original, as are the self-imposed constraints, namely to restrict to only local rewrites on graphs, regardless their interpretation as lambda terms (or lack of such interpretation), restrict to only two algorithms of reduction (deterministic or random), the most simple possible.

\section{Introduction}

    {\em "Define a molecular computer as one molecule which transforms, by random chemical reactions mediated by a collection of enzymes, into a predictable other molecule, such that the output molecule can be conceived as the result of a computation encoded in the initial molecule. "}  (source: \cite{7})

       (Computation of the Ackermann(2,2)=7   with chemlambda \href{http://chorasimilarity.github.io/chemlambda-gui/dynamic/expo_ackermann_2_2.html}{link to js animation}. See further details about how is made. You can move the atoms by click and drag.)

    The notion of a molecular computer presented here is  a model of computation based on individual molecules in a random environment. The model proposed here satisfies the following requirements: (a) is Turing universal (thus in principle to allow to compute anything), (b) based on a minimal set of well chosen chemical reactions which happen randomly.

    Individual molecules  encode data and programs in the chemical composition. Computations are not encoded in the numbers of species in a soup (multi set) of more or less arbitrarily chosen molecules.This is different than models of computation like the chemical abstract machine, or those based on chemical reaction networks.

        {\em I believe we can build molecular computers. This page is a call for making them real, either by discovering them in living organisms, or by learning how to build them from scratch.  }

        The argumentation of this possibility goes like this:     
    
\paragraph{Step 1: virtual molecular computers are possible.} Based on a made up, very simple artificial chemistry called  \href{http://chorasimilarity.github.io/chemlambda-gui/index.html}{chemlambda}, this  is proved theoretically \cite{1}, \cite{2}, \cite{3}. It is shown that chemlambda is Turing universal, that we can use some of the chemlambda molecules to encode untyped lambda beta calculus terms (but there exist many other molecules which do not encode lambda terms), that the reduction rules of the pure, untyped, lambda beta calculus can be realized by the chemlambda graph rewrites (but the graph rewrites apply as well to molecules which don't represent lambda terms and they work in a more general sense than only as representations of untyped lambda beta reduction rules).

         Chemlambda works with molecules which are graphs made by elementary constituents ("atoms") and links ("bonds") which enter in chemical reactions seen as local graph rewrites and interpreted as chemical reactions of the molecule with a set of invisible "enzymes", one enzyme per type of graph rewrite. Moreover, chemlambda does not use variables and does not rely on evaluations of terms. Instead of passing variables and evaluations, chemlambda uses a different strategy which is akin to  signal transduction.     
    
            To obtain a full model of computation one needs an algorithm concerning the application of graph rewrites. The algorithm is random and uses the fact that the graph rewrites are local (i.e. they involve only patterns with a small number of atoms and bonds, without any concern about the global structure of the graph-molecule). Chemlambda is therefore an asynchronous graph-rewriting automaton.

           There is a  \href{https://github.com/chorasimilarity/chemlambda-gui/blob/gh-pages/dynamic/README.md}{github repository} \cite{4} where this model is implemented and various   \href{http://chorasimilarity.github.io/chemlambda-gui/dynamic/demos.html}{demos}  are available \cite{9}, which show some features of the model: classical examples of  lambda calculus computations,  demos  about  artificial life forms called "chemlambda quines".      (How to use the repository:  the active branch of the repository is available at  \href{https://github.com/chorasimilarity/chemlambda-gui/tree/gh-pages/dynamic}{this link}. Follow the instruction from  \cite{4}.

       As an example, \href{http://chorasimilarity.github.io/chemlambda-gui/dynamic/sheet_copy.html}{this javascript animation} is the output of the computation which uses the molecule  \href{https://github.com/chorasimilarity/chemlambda-gui/blob/gh-pages/dynamic/mol/sheet_copy.mol}{sheet\_copy.mol}. It is produced by the scripts from the repository and you use the browser to see it.  It's an example of a computation which is not about lambda calculus, but nevertheless it leads to a structure which is programmed in the initial simpler one. You can move the atoms by click and drag.

\paragraph{Step 2: from virtual chemical reactions to real ones.} The patterns which appear in the chemlambda graph rewrites are so simple that it is highly possible that there exist real correspondents of "atoms", "ports", "bonds" and "enzymes", perhaps as simple real chemical molecules, which enter in  real chemical reactions which mimic the graph rewrites of chemlambda.  The problem is to identify those patterns in reality, by a combination of database searches and experiments which make this project ideal for an open source bio-hacking project, for example.

\section{Chemlambda} 
Chemlambda s a graph-rewriting formalism with some of the moves (graph-rewrites) inspired from lambda calculus, like for example the BETA move, which is a variant of the Wadsworth-Lamping treatment of the beta move on syntactic trees of lambda terms. But there is a difference: in chemlambda there are no correct molecules (graphs).    

   Another resemblance is with the  \href{http://tuvalu.santafe.edu/~walter/AlChemy/Statement/organization.html}{Alchemy of Fontana and Buss}. They also propose to look at lambda calculus as an artificial chemistry. The main difference is that while in  chemlambda application and abstraction are treated on the same par, as atoms in the molecule, in Alchemy application is seen as a chemical reaction and abstraction is seen as a reactive locus in a molecule, i.e. the chemical reactions in Alchemy are  of the type 
$$A + B  \, \longleftrightarrow \,  AB$$ 
 where A, B are lambda terms and AB is A applied to B.  In the work of   \href{https://www.tbi.univie.ac.at/~xtof/papers.html}{Christoph Flamm} the chemical reactions are of the type 
$$ A + B  \, \longleftrightarrow ,  C + D$$
 understood as a rewrite of the unconnected pattern obtained from the union of A and B. In chemlambda the rewrites model a chemical reaction 
$$A + Enzyme  \, \longleftrightarrow \,  B + Enzyme$$
 where Enzyme is associated to the rewrite.

\paragraph{Chemlambda molecules.} The molecules of chemlambda are graphs made by 9  elementary constituents, called A, L, FI, FO, FOE, T, Arrow, FRIN, FROUT. Every elementary  constituent is made by a main atom, represented in the demos as a big colored circle, and some smaller atoms, called ports.  A, L, FI, FO, FOE have 3 ports each, Arrow has two ports, and T, FRIN and FROUT have one port each. They are linked by bonds, which are of two kinds: inner, between a main atom and its ports, and outer, between ports.  Each port has a pair of types, one from the family (left, right, middle) and the other one from the family (in,out). The outer bonds are permitted only between a pair of ports, one with type in, the other with type out.

        We use the "mol" notation for a chemlambda molecule. The programs from the github repository work with files with the extension .mol as inputs. Thus a molecule is represented by a .mol file. The visualisations  represent a molecule as made by coloured circles of various radii (there is a radius "main\_const" for the main atoms and the types of ports  "left", "right" and "middle" are represented by radii of circles with the same names). 
 The colour of the main atoms comes  from the type of the atom (A, L, FI, ...) and the colour of the ports encode the type "in" or "out".    

     Therefore there is a one-to-one correspondence between the mol notation and what you see. Everything in the visualizations  is coded by color, radius of the circle and, for links, there are inner (thicker) and outer links. For reading convenience I shall use (a given pallete of) these colours in the explanations.

    The colours are:  "in\_col", "out\_col", "red\_col", "green\_col",  "arrow\_col" and "term\_col".  The colour palette does not really matter, of course, for the chemistry (see  \href{http://imar.ro/~mbuliga/chemlambda_story.html}{this older introduction about the molecules} with a different palette).      

    Here is the list of nodes with the types of their ports, and the way they are written in the mol file.  The colours are as in the visualisations convention. Recall that the mol file notation does not need colours, these are only for visualization purposes.
\begin{enumerate}
\item[-] (L          middle.in        left.out        right.out), is inspired from the lambda abstraction.    
\item[-] (A          left.in        right.in         middle.out), is inspired from application operation.   
\item[-] (FI          left.in        right.in         middle.out), has no correspondent in lambda calculus, name means "fan-in".   
\item[-] (FO          middle.in        left.out        right.out), has no correspondent in lambda calculus, name means "fan-out".    
\item[-] (FOE          middle.in        left.out        right.out), has no correspondent in lambda calculus, name means "extra fan-out".  Mind that in chamlambda there are two kinds of fan-out nodes!   
\item[-] (Arrow          middle.in        middle.out), has no correspondent in lambda calculus, is a simple solution for application of some of the moves. Is inspired from Louis Kauffman proposal to use commutative polynomials for the graphical elements (where the variables of the polynomials play the roles of the ports). Louis Kauffman used on July 3rd 2014  the Mathematica symbolic reduction  for polynomial commutative algebra in order to try to reduce the fixed point combinator in graphic lambda calculus, thus making the first attempt of a visualisation program for chemlambda.    
\item[-] (T          middle.in), called "termination" node, is used  in relation to lambda calculus in place of variables x appearing in terms $\lambda x.A$, where A does not contain x.    
\item[-] (FRIN          middle.out), is a  "free in" node, used  for edges with a free source. (Thus, if your mol file has a port variable "a" which appears only once, in a port of type in, then the main script adds "FRIN a" to the mol file.)     
\item[-] (FROUT          middle.in), is a  "free out" node, used  for edges with a free target. (Thus, if your mol file has a port variable "a" which appears only once, in a port of type out, then the main script adds "FROUT a" to the mol file.)     
\end{enumerate}

    Remark that the colours and the relative size of the circles are enough to encode all the information needed in the formalism. For example the colours of nodes and ports allows to discern which is    (L          1         2         3)     and which is    (FI          1         2          3).

\paragraph{Chemlambda graph rewrites.} A shorter name is "moves".  They should be seen as chemical reactions involving a small part of the molecule and an invisible enzyme, one enzyme type per move.    
    
    In the following you see the moves written in the mol convention, and also a visualization of the moves. For each move you can CLICK on the move name to see what is happening. You can move the atoms by click and drag.    

 \begin{enumerate}
\item[$\bullet$]  BETA and FAN-IN family 
 \begin{enumerate}
\item[-] BETA (or A-L) \href{http://chorasimilarity.github.io/chemlambda-gui/dynamic/expo_beta.html}{(link to js animation)}

$$\begin{array}{llll}
    L   &    1   &     2    &     c  \\
    A   &    c   &     4    &     3      
\end{array} \, \longleftrightarrow \, 
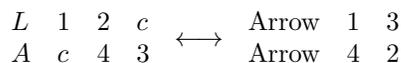
\begin{array}{llll}
    \mbox{Arrow}     &     1   &      3  & \\
    \mbox{Arrow}     &     4   &      2  & 
\end{array}  
$$
\item[-] FAN-IN (or FI-FOE) \href{http://chorasimilarity.github.io/chemlambda-gui/dynamic/expo_fanin.html}{(link to js animation)}

$$ \begin{array}{llll}
   FI    &      1   &     4     &   c   \\
   FOE   &      c   &     2     &   3      
\end{array} \,  \longleftrightarrow \,      
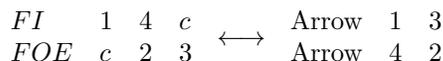
\begin{array}{llll}
 \mbox{Arrow}    &     1   &     3   & \\ 
 \mbox{Arrow}    &     4   &     2   & 
\end{array}
$$
        
\end{enumerate}

\item[$\bullet$] DIST family: 
\begin{enumerate}
\item[-] FO-FOE \href{http://chorasimilarity.github.io/chemlambda-gui/dynamic/expo_fofoe.html}{(link to js animation)}

$$ \begin{array}{llll}
  FO    &    1     &   2    &   c  \\
  FOE   &    c     &   3    &   4   
\end{array}  \, \longleftrightarrow \,   
\begin{array}{llll}
 FI      &   j    &    i    &    2   \\
 FO      &   k    &    i    &    3   \\
 FO      &   l    &    j    &    4   \\
 FOE     &   1    &    k    &    l    
\end{array}
$$

\item[-] FI-FO \href{http://chorasimilarity.github.io/chemlambda-gui/dynamic/expo_fifo.html}{(link to js animation)}

$$ \begin{array}{llll}
   FI    &      1   &      4      &    c   \\
   FO    &      c   &      2      &    3      
\end{array} \, \longleftrightarrow \, 
 \begin{array}{llll}
   FO    &     1     &    i   &     j    \\
   FI    &     i     &    k   &     2    \\
   FI    &     j     &    l   &     3    \\
   FO    &     4     &    k   &     l   
\end{array}
$$

\item[-] L-FOE  \href{http://chorasimilarity.github.io/chemlambda-gui/dynamic/expo_lfoe.html}{(link to js animation)} 

$$ \begin{array}{llll}
     L     &     1    &   2    &     c     \\
     FOE   &     c    &   3    &    4     
\end{array} \, \longleftrightarrow \, 
\begin{array}{llll}
    FI    &  j    &    i   &    2  \\
    L     &  k    &    i   &    3  \\
    L     &  l    &    j   &    4  \\
    FOE   &  1    &    k   &    l    
\end{array}
$$

\item[-] L-FO \href{http://chorasimilarity.github.io/chemlambda-gui/dynamic/expo_lfo.html}{(link to js animation)}

$$ \begin{array}{llll}
    L   &    1    &    2    &    c   \\
    FO  &    c    &    3    &    4   \\
\end{array} \, \longleftrightarrow \, 
 \begin{array}{llll}
    FI  &    j    &    i    &    2    \\ 
    L   &    k    &    i    &    3    \\
    L   &    l    &    j    &    4    \\ 
    FOE &    1    &    k    &    l      
\end{array} 
$$

\item[-] A-FOE \href{http://chorasimilarity.github.io/chemlambda-gui/dynamic/expo_afoe.html}{(link to js animation)}

$$ \begin{array}{llll}
     A   &    1    &    4     &    c    \\
     FOE &    c    &    2     &    3      
\end{array} \, \longleftrightarrow \, 
 \begin{array}{llll}
     FOE &    1    &    i     &   j     \\
     A   &    i    &    k     &   2     \\
     A   &    j    &    l     &   3     \\
     FOE &    4    &    k     &   l   
\end{array}
$$

\item[-] A-FO  \href{http://chorasimilarity.github.io/chemlambda-gui/dynamic/expo_afo.html}{(link to js animation)} 

$$ \begin{array}{llll}
      A   &   1     &    4     &   c     \\
      FO  &   c     &    2     &   3      
\end{array} \, \longleftrightarrow \, 
\begin{array}{llll} 
      FOE &   1      &   i     &   j     \\
      A   &   i      &   k     &   2     \\
      A   &   j      &   l     &   3     \\
      FOE &   4      &   k     &   l   
\end{array}
$$

\end{enumerate}

\item[$\bullet$] PRUNING family 
\begin{enumerate}
\item[-] A-T and FI-T, \href{http://chorasimilarity.github.io/chemlambda-gui/dynamic/expo_atfit.html}{(link to js animation)}  
     
$$ \begin{array}{llll}
       A  &    1      &   2      &    3    \\
       T  &    3      &          &         
\end{array} \, \longleftrightarrow \, 
\begin{array}{llll} 
       T  &    1      &          &          \\
       T  &    2      &          &    
\end{array}
$$

$$ \begin{array}{llll}
       FI &    1      &   2      &    3    \\
       T  &    3      &          &         
\end{array} \, \longleftrightarrow \, 
\begin{array}{llll} 
       T  &    1      &          &          \\
       T  &    2      &          &    
\end{array}
$$

\item[-] L-T     \href{http://chorasimilarity.github.io/chemlambda-gui/dynamic/expo_lt.html}{(link to js animation)}
   
$$ \begin{array}{llll}  
       L  &     1      &  2       &   3     \\
       T  &     3      &          &    
\end{array} \, \longleftrightarrow \, 
\begin{array}{llll}  
     T    &     1      &          &         \\
     T    &     c      &          &         \\
     FRIN &     c   
\end{array}
$$

\item[-] FO2-T, FOE2-T, \href{http://chorasimilarity.github.io/chemlambda-gui/dynamic/expo_fo2tfoe2t.html}{(link to js animation)}
 
$$ \begin{array}{llll}
     FO   &      1      &   2      &   3     \\
     T    &      2      &          &     
\end{array} \, \longleftrightarrow \, 
\begin{array}{llll}
     \mbox{Arrow}  &    1      &   3      &         
\end{array}
$$

$$ \begin{array}{llll}
     FOE  &      1      &   2      &   3     \\
     T    &      2      &          &     
\end{array} \, \longleftrightarrow \, 
\begin{array}{llll}
     \mbox{Arrow}  &    1      &   3      &         
\end{array}
$$

\item[-] FO3-T, FOE3-T, \href{http://chorasimilarity.github.io/chemlambda-gui/dynamic/expo_fo3tfoe3t.html}{(link to js animation)}

$$ \begin{array}{llll}    
    FO    &      1       &   2      &  3     \\
    T     &      3       &          &  
\end{array}    \, \longleftrightarrow \, 
\begin{array}{llll}
    \mbox{Arrow} &      1       &   2      &        
\end{array}
$$

$$ \begin{array}{llll}    
    FOE   &      1       &   2      &  3     \\
    T     &      3       &          &  
\end{array}    \, \longleftrightarrow \, 
\begin{array}{llll}
    \mbox{Arrow} &      1       &   2      &        
\end{array}
$$

\end{enumerate}

\item[$\bullet$] COMB move  \href{http://chorasimilarity.github.io/chemlambda-gui/dynamic/expo_arrow.html}{(link to js animation)} and COMB cycle. For the COMB move "M  1" denotes any node "M" which has (possibly among others) a out port "1". 

$$ \begin{array}{llll}   
     M   &    1     &      &   \\
     \mbox{Arrow}    &    1     &   2  & 
\end{array}   \, \longleftrightarrow \, 
\begin{array}{llll}
     M   &    2     &      &
\end{array}
$$

The cycle is a rapid application of COMB moves, whenever there are Arrow elements with the "in" port connected to a port which does not belong to another Arrow.     Exemplified here \href{http://chorasimilarity.github.io/chemlambda-gui/dynamic/expo_arrow_cycle.html}{(link to js animation)}  with a BETA move followed by a COMB cycle which eliminates all the Arrow elements, excepting the one which has its in port connected to its out port (so connected to another Arrow port).

\end{enumerate}

\paragraph{The reduction algorithm.}     One needs to add to chemlambda a reduction model (which says basically how to apply the moves and what to do if two or more moves attempt to modify the same node of the graph). Any reduction model will do, with the conditions to be:
\begin{enumerate} 
 \item[-]  local (the algorithm for deciding which move to apply or which solves conflicts has to be formulated so that it works with an input consisting of a graph with at most N nodes and links, with N fixed a priori)    
\item[-]  geometrical (the result of the algorithm is the same regardless of the internal numbering, ordering or naming of the variables from the input)   
\item[-]    it has to be asynchronous (because of locality in time as well) and decentralized, in the sense that the "local" constraint applies also in space, so the algorithm running in one place has to be able to solve moves and conflicts on its piece of graph based on an a priori bounded number N of graph nodes, links and messages exchanged with other places.       
\end{enumerate}

   In the demos \cite{9} there is used the most simple algorithm, called for this reason the "stupid" one, in order to show that even so the results are interesting. This algorithm may turn out to be less stupid than I thought, in particular because it led to this proposal of a  molecular computer.

   The algorithm takes as input an initial graph, then there is a cycle where, at each step (deterministically or randomly):
\begin{enumerate}
\item[-]  are identified the pieces of graph where the moves apply   
\item[-]  moves are applied: in the deterministic version, according to a priority of moves (i.e. in this case first identify the patterns for the move FO-FOE, then block the nodes which involve that move, then look further for the other moves DIST which involve free nodes, then block the nodes involved, then look for the BETA and FAN-IN moves, block the nodes, then look for PRUNING moves, block the nodes);  in the random version each move has an asociated random coin flip, with an weight. The coin is flipped for each proposed move and the move is applied or not, accordingly.   
\item[-] at the end of the step there is a COMB cycle, which search for the application of all COMB moves (which eliminate the Arrow elements) until there is no possibility of applying further COMB moves (which does not mean that all Arrow elements are eliminated)      
\end{enumerate}

   The effect of this  algorithm is equivalent, in case we start from a graph which represents a lambda term \cite{1} section 3, to a massively parallel reduction of the term, but one not involving any variable passing, nor evaluation steps.   

   On top of this, if we restrict to graphs coming from lambda terms, one has to take care that: (a) there is no eta reduction, it's pure lambda beta, so "functional" is not appropriate to use because functions need eta reduction (b) even if the initial and final graphs (if there is a final graph) correspond to lambda terms, typically the intermediary ones do not, see for example  \href{http://chorasimilarity.github.io/chemlambda-gui/dynamic/pred_2_bw.html}{the bw version of the predecessor reduction}.

   Each move is seen as an interaction between a (part of a) molecule-graph with an invisible enzyme, one enzyme per move. One can make the enzymes visible, even use them as a resource, introduce moves between enzymes, making the whole model conservative, or even transform the enzymes into chemlambda graphs as well, by a sort of bootstrapping.  [Added: in \cite{11} is presented a conservative (with respect to nodes and arrows), without enzymes, equivalent graph rewrite formalism for chemlambda.]


\begin{thebibliography}{99}

\bibitem{13} Berry, G\' erard, Boudol, G\' erard, The chemical abstract machine (1992),Theoretical Computer Science {\bf 96}, 1, p. 217 - 248

\bibitem{12} Lafont,  Yves,  Interaction  nets  (1990).  Proceedings  of  the  17th  ACMSIGPLAN-SIGACT  Symposium  on  Principles  of  Programming  Lan-guages. ACM: 95108

\bibitem{14} W. Fontana and L. W. Buss, What would be conserved if `the tape were played twice'?
Proc. Natl. Acad. Sci. USA, 91, 757-761 (1994)

\bibitem{15} W. Fontana and L. W. Buss
'The Arrival of the Fittest': Toward a Theory of Biological Organization
Bull. Math. Biol., 56, 1-64 (1994)

\bibitem{16} Martin Mann, Heinz Ekker, Christoph Flamm 
The Graph Grammar Library - a generic framework for chemical graph rewrite systems. In: Proceedings of the International Conference on Model Transformation (ICMT 2013), D. Varro and K. Duddy and G. Kappel (eds), Springer-Verlag Berlin Heidelberg, LNCS 7909, pp. 52-53, (2013). \href{http://dx.doi.org/10.1007/978-3-642-38883-5_5}{(doi)}

\bibitem{17} King, J. , The Chemlambda-hask repository \href{https://github.com/synergistics/chemlambda-hask}{Chemlambda-hask repository}

\bibitem{0} M. Buliga, Molecular computers (2015), \\
 \href{http://chorasimilarity.github.io/chemlambda-gui/dynamic/molecular.html}{http://chorasimilarity.github.io/chemlambda-gui/dynamic/molecular.html}

\bibitem{8} M. Buliga, The library of chemlambda molecules (2016) \\
\href{https://github.com/chorasimilarity/chemlambda-gui/tree/gh-pages/dynamic/mol}{https://github.com/chorasimilarity/chemlambda-gui/tree/gh-pages/dynamic/mol}


\bibitem{9} M. Buliga, The chemlambda project demos, \\
 \href{http://chorasimilarity.github.io/chemlambda-gui/index.html}{http://chorasimilarity.github.io/chemlambda-gui/index.html}

\bibitem{10} M. Buliga, The chemlambda collection of simulations (2017), Figshare \href{https://doi.org/10.6084/m9.figshare.4747390.v1}{(doi)}

\bibitem{11} M. Buliga, Chemlambda strings (2018), Figshare \href{https://doi.org/10.6084/m9.figshare.5751318.v1}{(doi)}

\bibitem{1} M. Buliga,  Graphic lambda calculus. Complex Systems 22, 4 (2013), 311-360      \href{http://www.complex-systems.com/abstracts/v22_i04_a01.html}{(journal)}  \href{http://arxiv.org/abs/1305.5786}{(arXiv)}

\bibitem{2}  M. Buliga,  Chemical concrete machine.  Figshare \href{http://dx.doi.org/10.6084/m9.figshare.830457}{(doi)},    \href{https://arxiv.org/abs/1309.6914}{(arXiv)}   

\bibitem{3} M. Buliga, L.H. Kauffman,  Chemlambda, universality and self-multiplication. Presented at ALIFE-14.    \href{https://www.mitpressjournals.org/doi/pdf/10.1162/978-0-262-32621-6-ch079}{MIT Press}    \href{https://arxiv.org/abs/1403.8046}{(arXiv)}
    
\bibitem{4} M. Buliga,  \href{https://github.com/chorasimilarity/chemlambda-gui/blob/gh-pages/dynamic/README.md}{Chemlambda repository}

\bibitem{5} M. Buliga, L.H. Kauffman, GLC actors, artificial chemical connectomes, topological issues and knots. \href{https://arxiv.org/abs/1312.4333}{(arXiv)}  

\bibitem{6}  M. Buliga,  Zipper logic. Figshare   \href{https://dx.doi.org/10.6084/m9.figshare.1032660}{(doi)}  \href{https://arxiv.org/abs/1405.6095}{(arXiv)}

\bibitem{7}    M. Buliga,  Build a molecular computer. Journal of Brief Ideas, 2015.      \href{https://beta.briefideas.org/ideas/ae644667a7badd7899570fb530016c7b}{(doi)} 



















\end{thebibliography}
\end{document}